\documentstyle[12pt,titlepage]{article} 
\title{Decay constants and mixing parameters in a relativistic model for 
$q\bar Q$ system}
\author{Mohammad R. Ahmady, Roberto R. Mendel and James D. Talman}          
\date{November, 1995}   
\def\_{\rule{.3em}{.15ex}}  
\def\lms{\Lambda_{\overline {MS}}}
\newcommand {\ix}{\hspace*{1em}}
\newcommand\bs{\char '134 }  
\begin{document}           
\begin{titlepage}
\begin{flushright}
 OCHA-PP-70\\
\end{flushright}
 \begin{center}
  \vspace{0.75in}
  {\bf {\LARGE Decay constants and mixing parameters in a relativistic model 
for 
  $q\bar Q$ system}} \\
  \vspace{0.75in}
 Mohammad R. Ahmady$^{\rm a}$, Roberto R. Mendel and James D. Talman$^{\rm b}$ 
 \vskip 1.0cm
$^{\rm a}${\small \it Department of Physics, Ochanomizu University, 1-1 
Otsuka 2, Bunkyo-ku, Tokyo 112, Japan} \\
$^{\rm b}${\small \it Department of Applied Mathematics,
  The University of Western Ontario,
  London, Ontario, Canada N6A 5B7}\\
  \vspace{1in}
  ABSTRACT \\
  \vspace{0.5in}
  \end{center}
  \begin{quotation}
  \noindent 
We extend our recent work, in which the Dirac equation with a ``(asymptotically 
free) Coulomb + (Lorentz scalar $\gamma_0\sigma r$) linear '' 
potential is used to obtain the light quark wavefunction for $q\bar Q$ 
mesons in the limit $m_Q\to \infty$, to estimate the decay constant $f_P$ 
and the mixing parameter $B$ of the pseudoscalar mesons.  We compare our 
results for the evolution of $f_P$ and $B$ with the meson mass $M_P$ to 
the 
non-relativistic formulas for these quantities and show that there is a 
significant correction in the subasymptotic region.  For $\sigma 
=0.14{{\rm ~GeV}}^{-2}$ and $\lms =0.240{\rm ~GeV}$ we obtain: $f_D 
=0.371\; ,\; f_{D_s}=0.442\; ,\; f_B=0.301\; ,\; f_{B_s}=0.368 {\rm ~GeV}$ 
and $B_D=0.88\; ,\; 
B_{D_s}=0.89\; ,\; B_B=0.95\; ,\; B_{B_s}=0.96\; ,\;$ and $B_K=0.60$.


\end{quotation}
\end{titlepage}


\newcommand{\da}{\mbox{$\scriptscriptstyle \dag$}}
\newcommand{\lag}{\mbox{$\cal L$}}
\newcommand{\tr}{\mbox{\rm Tr\space}}
\newcommand{\fc}{\mbox{${\widetilde F}_\pi ^2$}}
\newcommand{\ns}{\textstyle}
\newcommand{\si}{\scriptstyle}

Recently, we presented our results for the form factor (Isgur-Wise 
function) which parametrizes the transition matrix elements of mesons 
containing a heavy quark \cite{amt}.  In our relativistic treatment of a 
heavy-light system, we assumed that, with the heavy quark fixed at the 
origin, the light quark wavefunction obeys a Dirac equation with 
a spherically symmetric potential.  This potential, an asymptotically free 
Coulomb term plus a linear confining term, reflects the QCD interactions 
both at short and long distances.  In this way, we obtained the shape and 
the slope at zero recoil of the form factor in the leading $1/m_Q$ order 
by fitting our model parameters to the experimental results on the 
semileptonic B decays.  

In this paper, we extend our results by 
calculating the decay constants $f_P$ and mixing parameters $B$ of the 
pseudoscalar $q\bar Q$ mesons in our model.  $f_P$ parametrizes the 
matrix element for the decay of a pseudoscalar meson through the 
axial-vector current $A^5_\mu$:
\begin{equation}
<0|A^5_\mu (x)|P(k)>=ik_\mu f_P e^{-ik\cdot x}\; ,
\end{equation}
where $k$ is the four-momentum of the meson.  On the other hand, $B$ is 
related to the matrix element for neutral-meson mixing which is 
conventionally written as
\begin{equation}
<\bar P|{(V_\mu -A_\mu^5)}^2|P>=\frac{4}{3}f_P^2M_PB\; ,
\end{equation}
where $V_\mu$ is the vector current and $M_P$ is the meson mass.  From a
phenomenological point of view, knowledge of these matrix elements is 
necessary for extracting important quantities such as the quark mixing matrix 
and CP violation within the Standard Model.  Experimentally $f_\pi =132 
{\rm MeV}$ and $f_K =167 {\rm MeV}$ are 
well-known.  However, for heavy mesons only a 
few data with large uncertainties are available.  Mark III sets an upper 
bound $f_D <290 {\rm MeV}$ \cite{adler} 
while WA75\cite{aoki}, CLEOII\cite{acosta} 
and BES\cite{bai} find 
$225\pm 45\pm 20\pm 41 \; ,\; 344\pm 37\pm 52$ and $430^{+150}_{-130}\pm 
40 {\rm MeV}$ for $f_{D_s}$ respectively.

There are various quark-model, QCD sum rules and lattice calculations of 
these parameters\cite{rev,ukqcd,rh1,rh2}.  A frequently used point of 
comparison is a scaling law for $f_P$ that is derived from a 
non-relativistic (NR) quark model:
\begin{equation}
f_P^{NR}(m_Q\to\infty )\propto \sqrt{\frac{1}{M_P}} \; ,
\end{equation}
where $m_Q$ is the mass of the heavy quark.  The mixing parameter is 
identically 1 in any NR quark model:
\begin{equation}
B^{NR}(m_Q)=1\; .
\end{equation}

In Refs. \cite{rh1,rh2}, it is shown that the combination of a relativistic 
dynamics and an asymptotically free Coulomb interaction for $q\bar Q$ 
system results in a significant deviation from the NR scaling law.  Our 
approach here is complementary to Refs. \cite{rh1,rh2} in the sense that 
we use a potential which is not only asymptotically free but also 
includes a linear confining term which determines the global behavior of 
the Dirac wavefunction.  At the same time, we use a saturation value for 
the strong coupling $\alpha_s^\infty =\alpha_s (r\to\infty )$ therefore 
avoiding unphysical pair creation effects. 

We start with the time-independent Dirac equation:
\begin{equation}
\left [ \vec\alpha .(-i\vec\bigtriangledown ) +
V_c (r) +c_0 +\gamma^0 (\sigma r
+m_q) \right ] \Psi (\vec r )=\epsilon_q \Psi (\vec r ) \; ,
\end{equation}
where $m_q$ is the light quark mass.  $V_c(r)$ is the asymptotically free 
Coulomb potential (tranforming as the zeroth component of a Lorentz 
four-vector):
 \begin{equation}
V_c(r)=-\frac {4}{3}\frac {\alpha_s(r)}{r}
\end{equation}
where $\alpha_s(r)$ is obtained in the leading log approximation and is 
parametrized as follows
\begin{equation}
\alpha_s(r)=\frac {2\pi}{(11-\frac {2N_F}{3})\ln [ \delta +\frac {\beta 
}{r}]} \end{equation}
The parameter $\delta$ defines the ``long distance'' saturation value for 
$\alpha
_s$.  We use $\delta =2.0$ which corresponds to 
$\alpha_s(r=\infty )\approx 1.0$.  The
parameter $\beta$ is related to the QCD scale $\Lambda_{\overline {MS}}$ by
$\beta ={(2.23\Lambda_{\overline
{MS}})}^{-1}$ for $N_F=3$ (we use $N_F=3$ throughout this paper).
We take $\Lambda_{\overline {MS}}\approx 0.240$ {\rm ~GeV}
, which
corresponds to $\beta =1.87 {{\rm ~GeV}}^{-1}$ in most of the calculations.

To describe the long distance behaviour, a linear term 
($\gamma_0\sigma r$) is introduced in the potential
which transforms as a Lorentz scalar.  This form is favored based on 
many theoretical and phenomenological arguments \cite{mpsy,mt,sg}.  For 
the parameter $\sigma$, we mainly use a value $\sigma =0.14$ which is 
favored by a recent lattice estimate \cite{blszw} and can also be 
extracted from the Regge slope using two-body generalization of 
Klein-Gordon equation \cite{mt,sg}.  However, we also present results for 
$\sigma =0.25$ and $0.18 {{\rm ~GeV}}^2$, as 
the former value is compatible with 
the experimentally available $0.45 {\rm ~GeV}$ 2P-1S splitting for $D$ and $D_s$ 
system and the latter is extracted from Regge slope data using string 
model \cite{perkins}.  As mentioned in the discussion of our results in 
Ref. \cite{amt}, the best fit of our model to the recent CLEO 1994 data 
analysis favors a smaller value for the parameter $\sigma$.  

The potential in the Dirac equation includes an additive constant  
$c_0$, which is clearly subleading both in the
short and long distance regimes.  The only role of this constant is
to define the absolute scale of the light quark energy $\epsilon_q$ which 
is identified with the ''inertia'' parameter often introduced in heavy 
quark effective theory:
\begin{equation}
\epsilon_q\equiv \bar \Lambda_q
=\lim_{m_Q\to\infty} (M_P-m_Q) \; .
\end{equation}
Therefore, only the difference $\bar 
\Lambda_q-c_0\approx\epsilon_q-c_0$ can be extracted from Eq. 
(5).  As we indicate later, in the heavy limit $m_Q\to\infty$, $f_P$ and 
$B$ are not very sensitive to $\bar\Lambda_q$ (and in turn to $c_0$).

The ground state solution to Eq.(5) has the form: 
\begin{equation}
\psi (\vec r)= N\left (\begin{array}{l}\; \; \chi (r)\\ 
-i\sigma .\hat r \phi (r)\end{array}\right )\; ,
\end{equation}
where $N$ is a normalization constant.  Figure 1 illustrates the 
functional behaviour of the normalized large component $\chi^n(=N\chi (r))$ 
and small 
component $\phi^n(=N\phi (r))$ of the wavefunction $\psi (\vec r)$ for the 
cases 
where $m_q=0$ (appropriate for $q\equiv u,d$) and $m_q=0.175 {\rm ~GeV}$ (for 
$q\equiv s$).  We observe that $\chi (r)$ is finite and $\phi (r)\to 0$ 
as \hbox{$r\to 0$}.  This, as noted in Refs. \cite{rh1,rh2}, is due to 
using a 
running $\alpha_s(r)$ in the Coulomb potential (Eq.(6)) rather than a 
fixed $\alpha$ which would result a singular wavefunction at the origin.  
On the other hand, since our model incorporates the long distance 
behaviour of QCD interactions (through linear confining potential) and at 
the same time avoids a singularity in the Coulomb potential (by 
introducing a saturation value for $\alpha_s(r)$), the resulting 
wavefunction is physically meaningful for the whole range of $r$.  This 
is our main improvement over previous works where a Dirac equation along 
with the leading-log Coulomb potential has been applied to heavy-light 
mesons \cite{rh1,rh2}.  The functions $\chi (r)$, $\phi (r)$ and the 
normalization constant $N$ are independent of $m_Q$ in our leading order 
(in $1/m_Q$) approach where we assume that the heavy quark is fixed at 
the origin.  However, as it is illustrated in Figure 1, the $SU(3)_F$ 
symmetry breaking strange quark mass $m_s=0.175 {\rm ~GeV}$ results in about 
$20\%$ larger value for $N$ ($\chi (r)$ is independent of $m_q$ as $r\to 
0$).  The normalization constant $N$ depends on the global behaviour of 
the potential, i.e. $\beta$ and $\sigma$ (see Eqs.(5), (6) and (7)).  We 
mainly use $\beta =1.87$ (corresponding to $\lms 
=0.240 {\rm ~GeV}$) and $\sigma 
=0.14 {{\rm ~GeV}}^{-1}$ (see the discussion following Eq. (7)), however, 
we do vary these parameters to investigate the sensitivity of our results.

We now proceed to compute the decay constant $f_P$ and mixing parameter 
$B$ in our model.  The decay constant is given by the overlap integral 
\cite{dj}
\begin{equation}
f_P^2(m_Q)=\frac{12}{M_P}N^2\int d^3\vec r {\vert \Psi_Q(\vec 
r)\vert}^2\chi^2(r)\; ,
\end{equation}
where 12 is a color-spin-flavour coefficient and the quantum-mechanical 
fluctuations in the position of the heavy quark are assumed to be 
described by a non-relativistic wavefunction $\Psi_Q(\vec r)$.  
Considering that $\Psi_Q(\vec r)$ is ``spread`` over a distance 
$r_Q=O(1/m_Q)$, one can reduce (10) to the expression \cite{rh2}:
\begin{equation}
f_P(m_Q)=\sqrt{\frac{12}{M_P}}N\chi (r_Q)\; .
\end{equation}
At this point, we need to make a definite connection between $r_Q$ and 
the heavy quark mass $m_Q$.  It is assumed that $r_Q=\kappa /m_Q$, where
 $\kappa \ge 1$ is expected on physical grounds.  We obtain a value for 
$\kappa$ by 
extrapolating the decay constant formula (11) to K-meson system where 
$f_K=167 {\rm MeV}$ is known experimentally.  This extrapolation, even though 
of uncertain reliability due to finite $m_Q$ effects, has been frequently 
made in purely NR models.  For $m_s^{\rm constituent}\approx M_K=0.495{\rm 
~GeV}$, we obtain $\kappa =1.67$ 
for $\lms =0.240$ and $\kappa = 1.73$ for $\lms = 0.200$.
In comparing 
our results for the decay constant with the NR scaling law, Eq. (3), we 
also include a finite renormalization factor suggested by Voloshin and 
Shifman \cite{vs} and Politzer and Wise \cite{pw}.  In Figure 2, 
comparison is made between the ''improved'' NR scaling law
\begin{equation}
f_P^{NR}(M_P)\propto \sqrt{\frac{1}{M_P}}{\left 
(\frac{1}{\tilde{\alpha}_s(M_P)}\right )}^\gamma\; ,
\end{equation}
and our relativistic results
\begin{equation}
f_P(M_P)\propto \sqrt{\frac{1}{M_P}} \chi \left ( 
\frac{\kappa}{M_P}\right ){\left (\frac{1}
{\tilde{\alpha_s}(M_P)}\right )}^\gamma\; ,
\end{equation}
where in the latter we have made the approximation $M_P\approx m_Q$ in 
the expression for $r_Q$.  One can justify this approximation in 
view of the uncertainty in the determination of $r_Q$ from the 
extrapolation to K-meson system.  Because of this assumption, $r_Q$ does not 
depend
on the inertia parameter $\bar \Lambda_q$ and therefore is also independent of
the constant $c_0$.  Also in Eq. (13), $\gamma =2/9$ for $N_F =3$ and 
$\tilde{\alpha}_s(M_P)=2\pi /9\ln(M_P/\lms )$ in the renormalization 
correction 
factor. In the Figure, the above functions are normalized to their value at 
a large mass scale ($\sim 60\lms$) where the physics (of $m_Q\to\infty$) 
is well understood.  Our results 
are presented for $m_q=0$ and $m_q=0.175 {\rm ~GeV}$.  The main feature that 
distinguishes the relativistic graph from the NR scaling law is the 
maximum at around D meson mass, as noted in Refs. \cite{rh1,rh2}.  We 
would like to emphasize again, that our improved model is free of 
unphysical behaviour for the whole range of $r$.  Therefore, here we 
rigorously confirm that the maximum in the scaling law is {\it physical} 
and is due entirely to the relativistic dynamics of the light quark at 
the short distances.  The position of this maximum is around $7.0\lms$ 
for $m_q=0$ (for $\lms =0.240 {\rm ~GeV}\; ,\; M_P^{\rm peak}\approx 1.68 {\rm
 ~GeV}$ 
which is somewhat below $M_D\approx 1.87 {\rm ~GeV}$) and around $8.1\lms$ 
for 
$m_q=0.175 {\rm ~GeV}$ (resulting $M_{P_s}^{\rm peak}\approx 1.94 {\rm 
~GeV}$ which is 
roughly the same as $M_{D_s}\approx 1.96 {\rm ~GeV}$).  From Figure 2, we 
also notice that, 
roughly speaking, from the B meson mass scale onward, $SU(3)_F$ is a good 
symmetry as far as the scaling of the decay constant is concerned.

The mixing parameter $B$ can also be written as an overlap integral 
\cite{stc}  
\begin{equation}
B(m_Q)=\frac{12}{f_P^2M_P}N^2\int d^3\vec r {\vert \Psi_Q(\vec 
r)\vert}^2\left [\chi^2(r)-\phi^2(r)\right ]\; ,
\end{equation}
which will be approximated by 
\begin{equation}
B(m_Q)=1-{\left [\frac{\phi (r_Q)}{\chi (r_Q)}\right ]}^2\; ,
\end{equation}
by the same argument that followed Eq.(10).  The non-relativistic scaling 
law for mixing parameter (Eq. (4)) is recovered for $m_Q\to\infty$ (i.e. 
$r_Q\to 0$) as $\phi (r\to 0)=0$ (see Fig. 1). 

Figure 3 illustrates the evolution of the mixing parameter $B$ with the 
meson mass $M_P$ (as in the case of the decay constant, $r_Q=\kappa /m_Q$ 
and we use the approximation $M_P\approx m_Q$) for $m_q=0$ and $m_q=0.175 
{\rm ~GeV}$.  We observe that in the subasymptotic region, deviation from the 
NR scaling law is $5\%$ and $12\%$ for the B and D meson mass scale 
respectively.  On the other hand, the $SU(3)_F$ symmetry breaking effects 
in the scaling, even for $M_D \stackrel {<}{\sim}M_P\stackrel 
{<}{\sim}M_B$, is negligible.

In Table 1, we present our numerical predictions for the decay constant 
and mixing parameter of D, ${\rm D}_s$, B, ${\rm B}_s$ mesons plus an 
estimate of $B_K$, the mixing parameter for the K-meson.  
To test the 
sensitivity of these predictions, we vary the model parameters $\beta$ 
and $\sigma$.  For example, changing $\lms$ 
from $0.240 {\rm ~GeV}$ ($\beta =1.87 
{\rm ~GeV}^{-1}$) to $0.200 {\rm ~GeV}$ ($\beta 
=2.24 {\rm ~GeV}^{-1}$), results in a very small 
change (ranging from $1\%$ to $4\%$) in $f_P$ and $B$.  
The ratios 
$f_{B_s}/f_B=1.22$ and $f_{D_s}/f_D=1.18$ (for $\sigma =0.14{\rm 
~GeV}^{-1}$) are independent of $\lms$.  On the other 
hand, the sensitivity to the linear potential coefficient is much more 
significant.  Besides $\sigma =0.14 
{\rm ~GeV}^{-1}$, the decay constants and 
mixing parameters are also estimated for 
$\sigma =0.18$, $0.25 {\rm ~GeV}^{-1}$ (see the paragraph on the linear 
potential following Eq. (7)).  An increase of $30\%$ to $40\%$ in the decay 
constants is observed for increasing $\sigma$ from $0.14$ to $0.25 
{\rm ~GeV}^{-1}$.  However, the mixing parameters are 
far less sensitive to this model parameter.

The results obtained for $f_D$ all exceed the upper bound of 0.290 of Ref. 2.
However, the finite renormalization factor is uncertain at the lower 
mass of the D meson; if this factor, which is 1.27, is not included, 
$f_D$ is 0.292 for $\sigma=0.14$.  The results may therefore not be  
incompatible with Ref. 2 at the smaller value of $\sigma$.

A comparison between our predictions (Table 1) and other theoretical 
results \cite{rev} also favors a smaller value for $\sigma$.  Even though 
our results are larger than other theoretical predictions, for $\sigma 
=0.14{\rm ~GeV}^{-1}$ and taking into account the uncertainty factor 1.27 
(from the finite renormalization factor below D-meson mass scale) there are
 agreements with some lattice and potential model 
estimates.  However, QCD sum rules predictions are generally smaller than 
ours.  Our estimated $B_B$ and $B_{B_s}$ is on the low side of a recent 
lattice prediction by UKQCD\cite{ukqcd}.

In conclusion, we used a relativistic model with a phenomenological 
potential that accounts for QCD interactions at all length scales, to 
estimate the decay constant $f_P$ and mixing parameter $B$ of heavy-light 
mesons.  The evolution of $f_P$ and $B$ with the meson mass significantly 
deviates from the NR scaling law in the phenomenologically interesting 
subasymptotic region $M_D \stackrel {<}{\sim}M_P\stackrel 
{<}{\sim}M_B$.

\vskip 1.5cm
{\bf Acknowledgement}
\vskip 1.0cm
The authors would like to thank H. Trottier for useful discussions.  Work 
of M. A. is supported by the Japanese society for the Promotion of Science.

\newpage
\begin{flushleft}
\Large \bf
Table Captions
\end{flushleft}
{\Large \bf Table 1:}
The estimated decay constants (in {\rm ~GeV}) and mixing parameters for various 
choices of the model parameters.

\newpage
\begin{flushleft}
\Large \bf Figure Captions
\end{flushleft}
{\Large \bf Figure 1:}
The normalized large $\chi^n=N\chi (r)$ and small $\phi^n=N\phi 
(r)$ component of the wavefunction for $q\equiv u,d$ ($m_q=0$) and $q\equiv 
s$ ($m_q=0.175{\rm ~GeV}$). \\ 

\vskip 1cm
\hskip -0.6cm
{\Large \bf Figure 2:}
The evolution of the decay constant with the meson mass ($m_q=0$ and 
$m_q=0.175 {\rm ~GeV}$) compared with the NR scaling law. \\ 
\vskip 1cm
\hskip -0.6cm
{\Large \bf Figure 3:}
The scaling of the mixing parameter $B$ with the meson mass.

\newpage
\setlength{\unitlength}{0.240900pt}
\ifx\plotpoint\undefined\newsavebox{\plotpoint}\fi
\sbox{\plotpoint}{\rule[-0.200pt]{0.400pt}{0.400pt}}%


\newpage
{\hskip 6cm \bf \Large Table 1} \\
{
\hspace*{-2cm}
\begin{tabular}{c|c|c|c}
\hline
\multicolumn{1}{|c|}{             } & 
\multicolumn{1}{c|}{$\sigma =0.25$} &
\multicolumn{1}{c|}{$\sigma =0.18$} &
\multicolumn{1}{c|}{$\sigma =0.14$} \cr
\hline
\multicolumn{1}{|c|}{
$\beta =1.87$} & 
\multicolumn{1}{c|}{
$\begin{array}{l} 
f_D=0.519  \; f_B=0.427  \\
f_{D_s}=0.587 \; f_{B_s}=0.496 \\
B_D=0.87 \; B_B=0.95 \\
B_{D_s}=0.89 \; B_{B_s}=0.95 \\
B_K=0.58
\end{array}$} &
\multicolumn{1}{c|}{
$\begin{array}{l} 
f_D=0.429  \; f_B=0.350  \\
f_{D_s}=0.497 \; f_{B_s}=0.417 \\
B_D=0.87 \; B_B=0.95 \\
B_{D_s}=0.89 \; B_{B_s}=0.95 \\
B_K=0.59
\end{array}$} &
\multicolumn{1}{c|}{
$\begin{array}{l} 
f_D=0.371  \; f_B=0.301  \\
f_{D_s}=0.442 \; f_{B_s}=0.368 \\
B_D=0.88 \; B_B=0.95 \\
B_{D_s}=0.89 \; B_{B_s}=0.96 \\
B_K=0.60 
\end{array}$} \cr
\hline
\multicolumn{1}{|c|}{
$\beta =2.24$} & 
\multicolumn{1}{c|}{
$\begin{array}{l} 
f_D=0.514  \; f_B=0.418  \\
f_{D_s}=0.580 \; f_{B_s}=0.485 \\
B_D=0.88 \; B_B=0.95 \\
B_{D_s}=0.89 \; B_{B_s}=0.96 \\
B_K=0.59 
\end{array}$} &
\multicolumn{1}{c|}{
$\begin{array}{l} 
f_D=0.424  \; f_B=0.341  \\
f_{D_s}=0.491 \; f_{B_s}=0.406 \\
B_D=0.88 \; B_B=0.95 \\
B_{D_s}=0.90 \; B_{B_s}=0.96 \\
B_K=0.60 
\end{array}$} &
\multicolumn{1}{c|}{
$\begin{array}{l} 
f_D=0.367  \; f_B=0.292  \\
f_{D_s}=0.434 \; f_{B_s}=0.356 \\
B_D=0.88 \; B_B=0.95 \\
B_{D_s}=0.90 \; B_{B_s}=0.96 \\
B_K=0.61 
\end{array}$} \cr
\hline
\end{tabular}
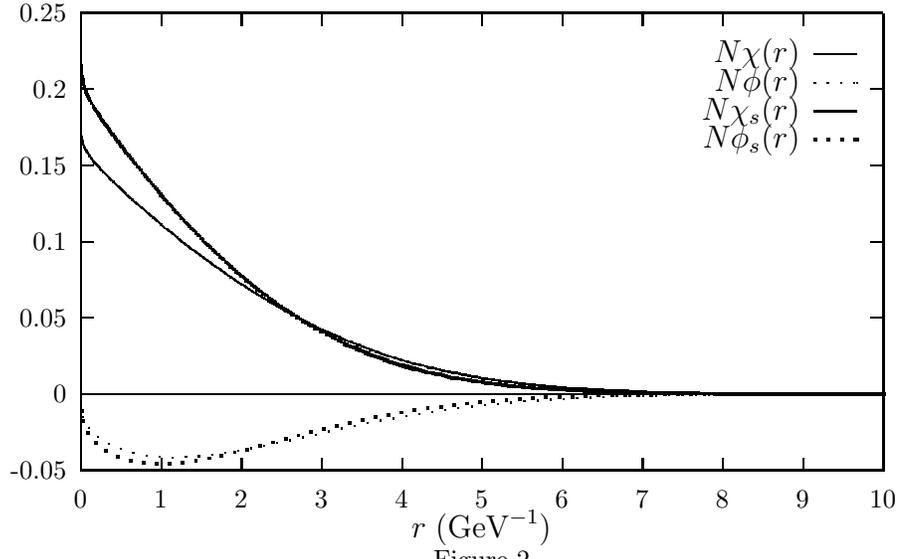
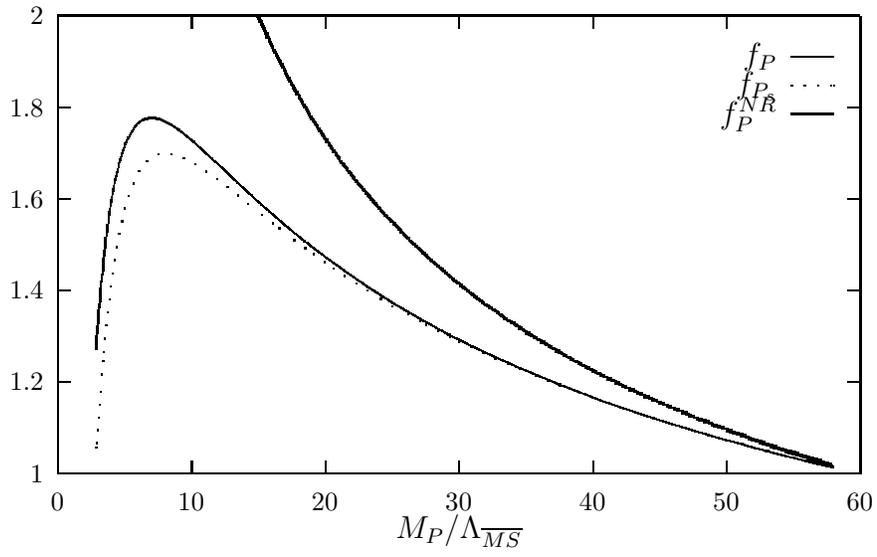
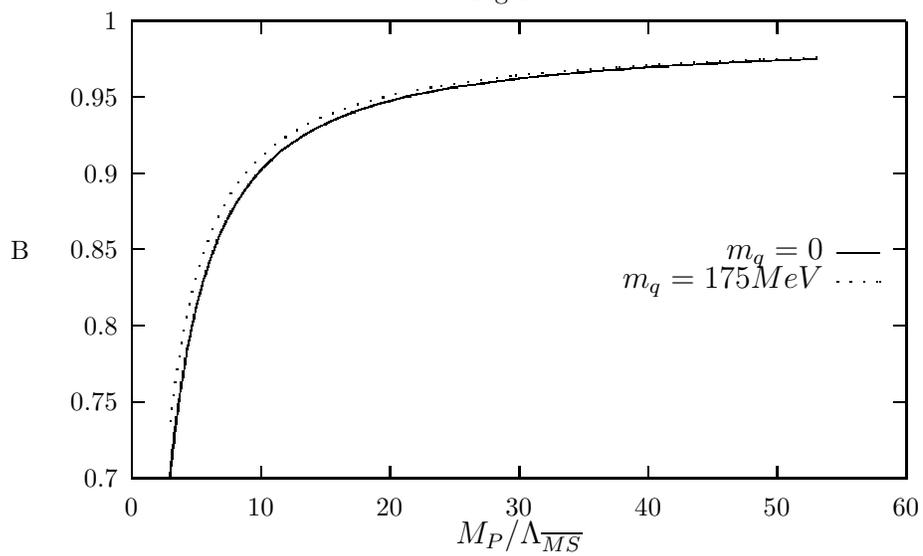
\end{document}